\documentstyle[12pt]{article}

\newcommand{\be}{\small\begin{equation}}
\newcommand{\ee}{\end{equation}\normalsize\vspace*{0.0ex}}

\newcommand{\bea}{\small\begin{eqnarray}}

\newcommand{\eea}{\end{eqnarray}\normalsize\vspace*{0.0ex}}

\newcommand{\bdm}{\small\begin{displaymath}}
\newcommand{\edm}{\end{displaymath}\normalsize\vspace*{0.0ex}}

\newcommand{\beas}{\small\begin{eqnarray*}}

\newcommand{\eeas}{\end{eqnarray*}\normalsize\vspace*{0.0ex}}

\newcommand{\n}{\noindent}

\begin{document}


\thispagestyle{empty}
\renewcommand{\thefootnote}{\fnsymbol{footnote}}

\setcounter{page}{0}
\begin{flushright} MPI-Ph/93-62\\
PSU/TH/130\\
hep-th/9309217\\
September 1993  \end{flushright}

\begin{center}
\vspace*{1.3cm}
{\Large\bf Instanton contributions\\[0.2cm]
 to the $\tau$ decay widths}\\
\vspace{2.0cm}
{\sc I.I. Balitsky\footnote{On leave of absence  from
St.Petersburg Nuclear Physics Institute, 188350 Gatchina,
Russia}} \\ \vspace*{0.7cm} {\it Physics Department, Penn State
University\\ 104 Davey Lab., University Park PA 16802,
USA}\\[1cm] {\sc M. Beneke} \hspace*{0.1cm}and\hspace*{0.2cm}
{\sc V.M. Braun$^*$} \\ \vspace*{0.7cm} {\it Max-Planck-Institut
f\"ur Physik\\ -- Werner-Heisenberg-Institut -- \\ D--80805
Munich (Fed. Rep. Germany)}\\[2.0cm]  {\bf Abstract}\\[0.3cm]
\end{center}
\parbox[t]{\textwidth}{
Contrary to some previous claims, we find a sizable instanton
contribution to the finite energy sum rule used to extract
the value of the strong coupling from the measured $\tau$ decay widths.
It is of the same order of magnitude as standard nonperturbative
corrections induced by vacuum quark and gluon condensates.
Our result indicates that there might be
no hierarchy of power corrections in finite energy
sum rules at the scale of $\tau$ mass. Therefore,
the standard nonperturbative corrections do not necessarily improve
the accuracy of the theoretical predicition, but
can rather be used  to estimate an
intrinsic accuracy of the pure perturbative calculation, which
turns out to be rather high on this evidence,
of order one percent.
}

\newpage


\n {\bf 1.}\hspace*{0.2cm}
In recent years the hadronic decay
widths of the $\tau$
lepton have attracted considerable attention
\cite{BRA92,DIB92,PICH92,ALE93}.
The  main interest originates from the observation that the
inclusive nature of the decay widths allows for a theoretical
prediction, comparison of which  with the data can fix the value
of the strong coupling with an accuracy competitive to its
determination at the $Z^0$ pole. A direct measurement of the
coupling at the $\tau$ scale might provide most conclusive
evidence for the perturbative running of the effective QCD
coupling. This is important, since the values of the coupling
extracted from deep inelastic scattering data and at the $Z^0$
pole could indicate a slower evolution compared to the QCD
renormalization group equations, a possibility that has already
triggered some activity towards alternatives to the standard QCD
evolution \cite{CLA92,JEZ93,ELL93}.

Due to momentum analyticity, all dynamical information on $\tau$
decays is contained in the two-point correlation functions of
flavour-changing vector and axial vector currents, which are
analyzed in the framework of the SVZ operator product expansion
\cite{SHI79}. In this approach the perturbative expansion of the
correlation functions is complemented by nonperturbative
corrections, which are proportional to the vacuum expectation
values of local operators build of quark and gluon fields
(condensates) and which are suppressed by increasing powers of the
$\tau$ mass $m_\tau \sim 1.8\,$GeV. The main evidence for the
applicability of standard SVZ expansions comes from QCD sum
rules, which are believed to work with the typical accuracy of
10--20\%. The problem is that a determination of
$\alpha_s(m_\tau)$ with, say, 10\% accuracy requires know\-led\-ge
of the decay widths at the one percent level. It has been found
that ``standard'' nonperturbative corrections to the $\tau$ decay
rate do not exceed a few percent \cite{BRA92}. This still
leaves open the possibility of nonperturbative effects that are
beyond the SVZ expansion, an issue that should be investigated
separately.

In the recent paper \cite{NAS93} Nason and Porrati study the
contribution to the total $\tau$ hadronic width from small
size instantons in the QCD vacuum. They find that even with the
highest value of $\alpha_s(m_\tau)$ allowed, the instanton
contributions never grow beyond $10^{-6}-10^{-5}$, and,
therefore, are completely negligible. This smallness can
essentially be traced to the well known fact \cite{THO76} that
the instanton density is proportional to the product of light
quark masses $m_u m_d m_s$ and instanton-induced transitions
vanish for strictly massless quarks. This is only true, however,
for one-instanton contributions in an empty (perturbative)
vacuum, which is far from a physical reality. As explained at
length in \cite{SHI80}, the density of small instantons of
size $\rho$ in the background field of large-scale vacuum
fluctuations is modified in such a way that the current quark
mass is substituted by an effective mass,

\be\label{effmass}
 m_q \rightarrow m_q -\frac{2}{3} \pi^2
\langle \bar q q \rangle \rho^2\,.
\ee

\n This effect of the vacuum ``medium'' is very large. To
obtain a rough estimate, one may replace the product of current
quark masses in the answer given in \cite{NAS93} by the product
of constituent masses $M_u \simeq M_d \sim 350\,$MeV,
$M_s \sim 500\,$ MeV. This yields an increase by a factor
$\sim 10^4$, and boosts the contribution obtained in
\cite{NAS93} into the range of percent corrections, which are
important. Thus, a more quantitative analysis of this effect is
mandatory, and this is the subject of the present paper. More
precisely, we calculate the instanton-induced contribution to
the relevant two-point correlation functions, which is related
in the framework of the operator product expansion to an
exponential correction to the coefficient function
in front of six-quark operators $(\bar{q} q)^3$. Since the
current quark masses on the right hand side of
(\ref{effmass}) are small compared to the second term, we shall
omit them altogether in the following.\\


\n {\bf 2.}\hspace*{0.2cm}
The correlation function of interest is the $T$-product of
two flavour-changing vector or axial vector currents,

\be
\Pi^{ud}_{\mu\nu}(q)\,=\,i\int\mbox{d}\Delta\,e^{iq\Delta}\,
\langle
0|T\left(j^\dagger_\mu(x)j_\nu(y)\right)|0\rangle\qquad\Delta
\equiv x-y\,,\ee
\be j_\mu(x)\,\equiv\,\bar{u}(x)\gamma_\mu\left\{
\begin{array}{c}1\\[-0.2cm] \gamma_5\end{array}\right\}
d(x)\,,\ee

\n of two quark fields, which we denote by $u$ and $d$, in a
theory with $n_f$ massless quarks. At large euclidean $q^2$ the
characteristic size $\rho$ of the instanton scales as $\rho^2
\sim q^{-2}\ll \Lambda^{-2}$, where $\Lambda$ is a typical QCD
scale parameter. Therefore an effective Lagrangian approach is
appropriate. To lowest order in $(\rho\Lambda)^2$ an instanton
of size $\rho$ (in the singular gauge) induces an effective $2
n_f$-quark vertex with the quark legs corresponding to the
instanton zero modes. Since these zero modes possess definite
chirality it is most convenient to use the two-component Weyl
spinor notation: the euclidean quark fields and $\gamma$-matrices
are written as

\be \label{notate}
q=\left(\begin{array}{c} i\chi^k_{\dot\alpha} \\ \psi^{k\alpha}
   \end{array}\right)\,,\hspace{0.5cm}
\bar q = (\bar\psi^{\dot\alpha}_k, i\bar\chi_{k\alpha})
\,,\hspace{0.5cm}
\gamma_\mu =\left(\begin{array}{cc}
0 & \bar\sigma_{\mu\dot\alpha\alpha} \\
\sigma_\mu^{\alpha\dot\alpha} & 0  \end{array}\right)\,,
\ee

\n
where $k=1,2,\ldots, N_c$
is a colour and $\alpha, \dot\alpha=1,2$ are spinor indices.
We use the notations
$\sigma_\mu^{\alpha\dot\alpha} = (1,-i\vec\sigma),
\bar\sigma_{\mu\dot\alpha\alpha} =(1,i\vec\sigma)$, where
$\vec\sigma$ are the Pauli matrices, and for vectors $v_\mu$
define $ v \equiv v_\mu \sigma_\mu,
\bar v \equiv v_\mu \bar\sigma_\mu $.
Whenever a mixture of spinor and colour indices takes place,
it is understood that spinor matrices act in the $2\times 2$
upper left corner of $N_c\times N_c $ colour matrices.
The
instanton vertex is then described by 't~Hooft's effective
Lagrangian

\be \label{efflag} L_\psi^I=(4\pi^2\rho_I^3)^{n_f} O_I\,,\quad
O_I=\prod_{i=1}^{n_f} (\bar{\chi}_i\varphi) (\bar{\kappa}
\psi_i)\,,\ee

\n where $\varphi^{k\alpha}=\epsilon^{\alpha\beta} U_{I\beta}^k$,
$\bar\kappa_{k\alpha}=-\epsilon_{\alpha\beta}(U_I^\dagger)^\beta_k$,
 $U_I$ is the $SU(N_c)$-matrix of
the instanton orientation and $\epsilon^{\alpha\beta}$
is the antisymmetric
tensor with $\epsilon^{12}=1$.
 Colour and spinor indices
will be suppressed, whenever no confusion can arise.

Let us recall the derivation of this expression.
 To find the coefficient multiplying the operator
$O_I$, one should compare the $2 n_f$-quark Green function
$\prod_{i=1}^{n_f}\psi_i(x_i)\bar{\chi}_i(y_i)$, evaluated in the
instanton background in the near mass shell limit
$x_i,y_i\rightarrow\infty$, with the result obtained from the
effective Lagrangian (\ref{efflag}). The $2 n_f$-quark amplitude
is simply obtained by substituting the zero modes for the quark
fields, i.e. equals
$\prod_{i=1}^{n_f}\kappa_0(x_k)\bar{\varphi}_0(y_k)$.
Explicit expressions for the zero modes for the instanton
with center at the origin are

\be \label{zeromodes}
\kappa_0(x)=\frac{\bar x \varphi }{2\pi^2x^4}
\frac{2\pi\rho_I^{3/2}}{\Pi_x^{3/2}}\,,\qquad
\bar{\varphi}_0(x)=\frac{\bar \kappa x}{2\pi^2x^4}
\frac{2\pi\rho_I^{3/2}}{\Pi_x^{3/2}}\,,\ee

\n where  $\Pi_x=1+\rho_I^2/x^2$. For the instanton with
center at $x_0$, the zero modes are obtained by the obvious
substitution $x\rightarrow x-x_0$.
 The large distance
behaviour of the zero modes coincides (in the singular gauge)
with that of the perturbative propagator up to the factors
$2\pi\rho^{3/2}_I\varphi$, $2\pi\rho^{3/2}_I\bar{\kappa}$,
respectively. Therefore the leading term of an expansion of the
zero modes in $\rho_I^2/x^2$ is indeed reproduced by an insertion
of the effective Lagrangian (\ref{efflag}). Subsequent terms in
this expansion correspond to the exchange of soft gluons, that
is, to higher dimensional operators in the effective Lagrangian
containing extra gluon fields.

Let us now return to the correlation functions. In two-component
notation

\be j_\mu(x)= \bar\psi_u(x)\bar\sigma_\mu\psi_d(x)
+\lambda \bar\chi_u(x)\sigma_\mu\chi_d(x)\,,
\qquad\lambda=\left\{\begin{array}{rl} -1
&\mbox{vector}\\[-0.2cm] 1 &\mbox{axial vector}\end{array}
\right.\,.\ee

\n In order to find the one-instanton contribution to the
coefficient function of the operator of lowest dimension,
$O_I$, one inserts the current product into a $2 n_f$-point
Green function, evaluates it in the instanton background and
takes the large distance asymptotics. The relevant diagrams for
all possible contractions of

\be \left[\prod_{i=1}^{n_f}
\psi(z_i)\bar{\chi}(z^\prime_i)\right]\,j_\mu^\dagger(x)
j_\nu(y)\qquad z_i,z_i^\prime\rightarrow\infty\ee

\n are shown in Fig.1, where the instanton is depicted as its
effective $2 n_f$-quark vertex. The solid line denotes the quark
zero modes and the dashed line represents the quark propagator
over the nonzero modes in the instanton background. Its explicit
form is known from \cite{BRO78}:

\bea \label{propagators}
S_I(x,y) \equiv   -i\langle \psi(x) \bar\psi(y)\rangle_I &=&
\frac{1}{\sqrt{\Pi_x \Pi_y}}
\Bigg\{
\frac{\Delta}{2\pi^2\Delta^4}\left(1-U_I P U_I^\dagger\right)
\\
&&\mbox{}+
\frac{\Delta}{2\pi^2\Delta^4}\left(1+
\rho_I^2\frac{U_I x\bar y U_I^\dagger}{x^2y^2}\right)
+\frac{\sigma_\mu \rho_I^2}{4\pi^2\Delta^2}
\frac{U_I x \bar \Delta \sigma_\mu \bar y  U_I^\dagger}
{x^2y^2(y^2+\rho_I^2)}\Bigg\}
\,,
\nonumber \\
\bar S_I(x,y) \equiv   i\langle \chi(x) \bar\chi(y)\rangle_I &=&
\frac{1}{\sqrt{\Pi_x \Pi_y}}
\Bigg\{ \frac{\bar\Delta}{2\pi^2\Delta^4}
\left(1-U_I P U_I^\dagger\right)
\nonumber\\
&&\mbox{}+
\frac{\bar\Delta}{2\pi^2\Delta^4}\left(1+
\rho_I^2\frac{U_I x\bar y U_I^\dagger}{x^2y^2}\right)
+\frac{\bar\sigma_\mu \rho_I^2}{4\pi^2\Delta^2}
\frac{U_I x \bar \sigma_\mu \Delta\bar y  U_I^\dagger}
{x^2y^2(x^2+\rho_I^2)}\Bigg\}
\,, \nonumber
\eea

\n where $P$ is the projector onto the $2\times2$ upper left
corner of the $N_c\times N_c$ matrix of the instanton orientation.
Just as in the case of the zero modes the near mass shell
asymptotics of these propagators coincides with the
perturbative propagators up to factors $1/\sqrt{\Pi}$:

\be\label{propasymptotics}
 S_I(x,z)\stackrel{z\rightarrow\infty}{\longrightarrow}
\frac{1}{\sqrt{\Pi_x}}\frac{-z}{2\pi^2 z^4}\qquad
\bar S_I(z,x)\stackrel{z\rightarrow\infty}{\longrightarrow}
\frac{1}{\sqrt{\Pi_x}}\frac{\bar{z}}{2\pi^2 z^4}
\ee

\n
Thus amputation of the external legs in the near
mass shell limit, denoted by a cross in the diagrams in Fig.1,
leaves the spinors $\kappa$ and $\bar{\varphi}$ for the zero
mode legs, see eq.(\ref{zeromodes}), and a factor $1/\sqrt{\Pi}$
for the propagator legs from (\ref{propasymptotics}). The leading
term in the operator product expansion of the current product in
the instanton background is then given by

\bea
&&\hspace*{-0.7cm}\int\!\mbox{d}\Delta\,e^{iq\Delta}\,
\langle j_\mu(x) j_\nu(y)\rangle^I_{(\bar\psi\psi)^{n_f}}\,=\,
\int\!\mbox{d}\Delta\,e^{iq\Delta}\int\frac{\mbox{{\footnotesize
d}}\rho}{\rho^5}\,d(\rho)\,\mbox{d} x_0\,\mbox{d} U\,
(4\pi^2\rho^3)^{n_f-1}\prod_{i\neq
u,d} \left[\!(\bar{\chi}_i\varphi)(\bar{\kappa}\psi_i)
\right]\nonumber\\
&&
\hspace*{-0.7cm}
\times\Bigg\{(4\pi^2\rho^3)\Big[\mbox{tr}
\left(\bar{\sigma}_\mu
S_I(x,y)\bar{\sigma}_\nu S_I(y,x)\right) + \mbox{tr} \left(
\sigma_\mu
\bar{S}_I(x,y)\sigma_\nu \bar{S}_I(y,x)\right)\Big]
(\bar{\chi}_u\varphi) (\bar{\kappa}\psi_u) (\bar{\chi}_d\varphi)
(\bar{\kappa}\psi_d)\nonumber\\
&&\hspace*{-0.7cm}
-\, 2\pi\rho^{3/2}\,\frac{1}{\sqrt{\Pi_x}}\,\Bigg[(\bar{\chi}_u
\varphi)
(\bar{\varphi}_0(y)\bar{\sigma}_\nu S_I(y,x)\bar{\sigma}_\mu\psi_u)
(\bar{\chi}_d\varphi) (\bar{\kappa}\psi_d)\\
&&\hspace*{-0.7cm}
+ \,(\bar{\chi}_u\varphi) (\bar{\kappa}\psi_u) (\bar{\chi}_d
\sigma_\mu \bar{S}_I(x,y)\sigma_\nu\bar{\kappa}_0(y)) (\bar{\kappa}
\psi_d)\Bigg] + (\,u\leftrightarrow d\,,\,\mu\leftrightarrow\nu\,,
\,x\leftrightarrow y\,)\nonumber\\
&&\hspace*{-0.7cm}
+\frac{1}{\sqrt{\Pi_x\Pi_y}}\,\Bigg[\Big\{(\bar{\chi}_u\varphi)
(\bar{\varphi}_0(x)\bar{\sigma}_\mu\psi_u) (\bar{\chi}_d\varphi)
(\bar{\varphi}_0(y)\bar{\sigma}_\nu\psi_d)
+ (\bar{\chi}_u\sigma_\nu\kappa_0(y)) (\bar{\kappa}\psi_u)
(\bar{\chi}_d\sigma_\mu\kappa_0(x)) (\bar{\kappa}\psi_d)\Big\}
\nonumber\\
&&\hspace*{-0.7cm}
+\lambda\,\Big\{(\bar{\chi}_u\sigma_\nu\kappa_0(y))
 (\bar{\varphi}_
0(x)\bar{\sigma}_\mu\psi_u) (\bar{\chi}_u\varphi) (\bar{\kappa}
\psi_u)
+(\bar{\chi}_d\sigma_\mu\kappa_0(x)) (\bar{\varphi}_0(y)\bar{
\sigma}_\nu\psi_d) (\bar{\chi}_d\varphi) (\bar{\kappa}\psi_d)
\Big\}\Bigg]\Bigg\}\,.\nonumber
\eea

\n Only the last term coming from the last diagram of Fig.1
differs by its sign for the vector and axial vector currents.
The integration over the instanton size is performed with the
instanton density \cite{BER79}

\be\label{instantondensity}
d(\rho)=\frac{c_1}{(N_c-1)! (N_c-2)!}\,e^{-N_c c_2+n_f
c_3}\,\left(\frac{2\pi}{\alpha(\rho)}\right)^{2 N_c}\,
e^{-2\pi/\alpha(\rho)}\,,\ee

\n where $c_1=0.466$ and the constants $c_2,c_3$ take the
values $c_2=1.54,c_3=0.153$ in the $\overline{MS}$ scheme. Now
it is only a matter of patience to integrate over the instanton
position and size and to take the Fourier transform.
A typical $\rho$-integral has the structure

\be\label{rhoint}
\int\limits_0^\infty \frac{\mbox{{\footnotesize d}}\rho}
{\rho^5}\,\rho^{2 B+1}\,\ln^E\frac{1}{\rho^2\Lambda^2}\,
K_1(\rho q)\,,\ee

\n where $K_1$ is a modified Bessel function and $B,E$ are some
numbers.
The logarithmic size-dependence stems
from three sources: from the two-loop running of $\alpha(\rho)$
in the exponent of the instanton density, from the preexponential
factor $\alpha(\rho)^{-2 N_c}$ and from the anomalous dimension
of the operators in the effective Lagrangian which are normalized
at the scale $\mu=1/\rho$.\footnote{
For consistency, we adjust the values of the one-loop and the
two-loop $\Lambda$ to reproduce the same value of $\alpha(m_\tau)$,
which is the only parameter appearing in the final formulas.}
 To perform this
integral, we first calculate a similar integral with the
logarithmic factor substituted by a power-like one
$(\rho\Lambda)^{-2 \epsilon}$,
and introduce the projection operator

\be \label{projector}
{\cal P}_E^\epsilon\, \{ F(\epsilon)\}
\equiv \frac{\Gamma(E+1)}{2\pi i} \int_C
\frac{\mbox{{\footnotesize d}}\epsilon}{\epsilon^{E+1}}
 F(\epsilon)
\,,\ee

\n where the contour $C$ wraps around the negative real axis.
For example, for integer $E$, or general $E$ and $\ln x>0$,
one has
${\cal P}_E^\epsilon\,\left\{x^\epsilon\right\}=\ln^E x$.
 For arbitrary real $E$
this operation produces an asymptotic series

\be \label{projectorexpansion}{\cal
P}_E^\epsilon\,\left\{\left(\frac{q^2}{\Lambda^2}\right)^
\epsilon Z(\epsilon)\right\}=\ln^E\left(\frac{q^2}
{\Lambda^2}\right)\sum_{n=0}^\infty\,\frac{\Gamma(E+1)}
{\Gamma(E+1-n)}\,\frac{z_n}{\ln^n (q^2/\Lambda^2)}\ee

\n with $Z(\epsilon)=\sum z_n\epsilon^n$ being
the result of the
$\rho$-integration. For natural $E$ the series terminates and
yields an exact answer for integrals of the type (\ref{rhoint}).
 For general real $E$ the
asymptotic
expansion in
(\ref{projectorexpansion}) yields a good approximation
provided $\ln (q^2/\Lambda^2)$ is sufficiently large,
and provided the
integral is dominated by $\rho<\Lambda^{-1}$.

Using the projection operator defined in (\ref{projector}),
we can write down the instanton contribution to the two-point
correlation function of (axial) vector currents in the background
of vacuum fermion fields in the compact form

\bea\label{ope}
\int\mbox{d}\Delta\,e^{iq\Delta}\,
\langle j_\mu(x) j_\nu(y)\rangle^I_{(\bar\psi\psi)^{n_f}} & = &
4^{B-2} (4\pi^2)^{n_f}\,\frac{c_1}{(N_c-1)! (N_c-2)!}\,e^{-N_c
c_2+n_f c_3}\,
\nonumber\\
&&\hspace*{-4cm} \times
\left(\frac{1}{q^2}\right)^{3 n_f/2}
\left(\frac{2\pi}{\alpha(q)}\right)^{2
N_c}  e^{-2\pi/\alpha(q)}\int\mbox{d}U\,
\prod_{i\not=u,d} (\bar{\chi}_i\varphi) (\bar{\kappa}\psi_i)\,
(-\beta_0\alpha(q))^E\nonumber\\
&&\hspace*{-4cm}\times
{\cal P}_E^\epsilon\,\Bigg\{e^{-\epsilon/(\beta_0\alpha(q))}
\,F(\epsilon) \Bigg[\left(\delta_{\mu\nu} q^2-q_\mu
q_\nu\right)\,\frac{2}{2 B-1-2 \epsilon}\,(\bar{\chi}_u\varphi)
(\bar{\kappa}\psi_u) (\bar{\chi}_d\varphi) (\bar{\kappa}\psi_d)
\nonumber\\[0.3cm]
&&\hspace*{-3cm} +\,(\bar{\chi}_u\varphi) (\bar{\kappa} [
q\bar{\sigma}_\mu-q_\mu]\psi_u) (\bar{\chi}_d\varphi)
(\bar{\kappa}[q\bar{\sigma}_\nu-q_\nu]\psi_d)\nonumber\\[0.3cm]
&&\hspace*{-3cm} +\,(\bar{\chi}_u
[\sigma_\nu\bar{q}-q_\nu]\varphi) (\bar{\kappa}\psi_u)
(\bar{\chi}_d[\sigma_\mu \bar{q} -q_\mu]\varphi)
(\bar{\kappa}\psi_d) \nonumber\\[0.3cm]
&&\hspace*{-3cm}
+\,\lambda\,(\bar{\chi}_u[\sigma_\nu \bar{q} -q_\nu]\varphi)
(\bar{\kappa}\psi_u) (\bar{\chi}_u\varphi) (\bar{\kappa}
[q\bar{\sigma}_\mu-q_\mu]\psi_u) \nonumber\\
&&\hspace*{-3cm}+\,\lambda\,(\bar{\chi}_d [\sigma_\mu \bar{q}
-q_\mu]\varphi) (\bar{\kappa}\psi_d) (\bar{\chi}_d\varphi)
(\bar{\kappa}[q\bar{\sigma}_\nu-q_\nu]\psi_d) \Bigg]\Bigg\}\,.
\eea

\n Here we defined

\be B=\frac{b+3 n_f}{2}\,,\quad b=-4\pi\beta_0\,,\quad E=2
N_c-\frac{2\pi\beta_1}{\beta_0}+\gamma\,,\ee
\be F(\epsilon)\equiv 4^{-\epsilon}\,\frac{\Gamma(B-\epsilon)
\Gamma(B-2-\epsilon) \Gamma(B-1-\epsilon)^2}{\Gamma(2
B-2-2\epsilon)}\,,\ee

\n where
$\beta_0,\beta_1$ are the first two coefficients of the
$\beta$-function, $\beta(\alpha)\equiv\mu^2\partial/\partial
\mu^2\,\alpha=\sum\beta_n\alpha^{n+2}$. The quantity
 $\gamma$ stands
schematically for the anomalous dimensions of the six-quark
operators -- that should be normalized at $q$ -- appearing in
the square brackets in eq.(\ref{ope}). An overall factor of two,
which accounts for an equal contribution in the background of an
antiinstanton, is included in (\ref{ope}).

A comment is in order, concerning
the integration over the instanton size. The
integration over $\rho$ contains a divergent term $\sim
(\delta_{\mu\nu} q^2-q_\mu q_\nu)\,1/q^4
\int\mbox{d}\rho\,\rho^{2 B-5}$. This term comes from large
distances and has to be identified with an instanton
contribution to the matrix element of the operator $\alpha
G_{\mu\nu} G_{\mu\nu}$. It will be neglected in the following.
The remaining terms are convergent and represent the instanton
contribution to the coefficient functions of the $2 n_f$-quark
operators shown in (\ref{ope}). The distinction between
contributions to the matrix elements and to the coefficient
functions is unambiguous. The relevant terms can easily be
distinguished by their $q^2$-dependence.

The answer in (\ref{ope}) has to be averaged
over the physical vacuum. For numerical estimation, we assume
factorization of the vacuum expectation values of the relevant
six-quark operators, which allows to express them in terms of the
quark condensate $\langle \bar q q \rangle$.
 Upon factorization, the second
and the third term in the square brackets in (\ref{ope}) vanish.
Since we are working in the chiral limit
$m_u=m_d=m_s=0$, we use the $SU(3)$-flavour
symmetry $\langle \bar{u} u\rangle=\langle\bar{d}d\rangle =
\langle \bar{s} s\rangle$, and take the numerical value
$\langle \bar{q} q\rangle=-(240\, MeV)^3 $ at the scale 1 GeV.
Extracting the Lorentz structure,
$\Pi^{ud}_{\mu\nu}(q)=(q_\mu q_\nu - \delta_{\mu\nu} q^2)
\,\Pi^{ud}(q^2)$, and taking the trivial integration over the
instanton orientation, we arrive at

\bea\label{instcorrelation}
\Pi^{ud}_{inst}(q^2) &=& -2\cdot
4^{B-2}\left(\frac{2\pi^2}{N_c}\right)^{n_f} (2\pi)^{2
N_c}\,\frac{c_1}{(N_c-1)! (N_c-2)!}\,e^{-N_c c_2+n_f c_3}
\nonumber\\
&&\times\prod_{i=1}^{n_f}\left[\frac{\langle\bar{q}_i q_i
\rangle(\mu)}{q^3}\alpha(\mu)^{\gamma/n_f}\right]\, e^{-2\pi/\alpha(q)}\,
\left(\frac{1}{\alpha(q)}\right)^{2 N_c+\gamma}
\nonumber\\
&&\times\,(-\beta_0\alpha(q))^E\, {\cal
P}_E^\epsilon\,\Bigg\{e^{-\epsilon/(\beta_0\alpha(q))}
\,F(\epsilon) \Bigg[\frac{1}{2 B-1-2\epsilon}-\lambda
\Bigg]\Bigg\}\,.\eea

\n For the anomalous dimensions of the six-quark operators
factorization amounts to taking $\gamma=4 n_f/b$, where $4/b$
is the anomalous dimension of the quark condensate.
However, since factorization is known not to be consistent with
the renormalization group,
we should rather allow $\gamma$ to vary in
a certain range, ascribing the variation of $\Pi_{inst}^{ud}$ to
the uncertainty inherent to
the factorization hypothesis.\footnote{
Strictly speaking, in presence of instantons, the renormalization
group equations for $(2n_f)$-quark operators are more complicated
and include mixing with the unity operator. This can be checked
by a direct calculation of the (anti)instanton contribution to
the vacuum expectation value $\langle(\bar q q)^{n_f}\rangle$,
which proves to contain a logarithm of the renormalization scale.
Taking into account this mixing corresponds to the calculation
of correlation functions in the instanton-antiinstanton background
in the spirit of \cite{BAL91}, but does not seem appropriate
for our present purposes since the size of the antiinstanton in
this calculation turns out to be unacceptably large.}

Corrections to eq.(\ref{instcorrelation}) come from (i)
higher-dimensional operators, suppressed by powers of
$(\rho\Lambda)^2$ and (ii) exchange of hard particles, suppressed
by powers of $\alpha(\rho_*)$, where $\rho_*$ is the average
size of instantons that contribute to the correlation functions.
{}From now on we shall discuss the case $n_f=N_c=3,q=m_\tau$,
which is the case of physical interest in $\tau$ decays, and
regard $\Pi^{ud}_{inst}$ as a function of $\alpha(m_\tau)$. Then
$b=B=9$ and $E=34/9$. To estimate the value of $\alpha(m_\tau)$,
at which the above mentioned corrections become of the same
order as the contribution kept in (\ref{instcorrelation}), we
should first find $\rho_*$. The distribution of instanton sizes
relevant to the term proportional to $\lambda$ in
(\ref{instcorrelation}) is given by the integral

\be\int\frac{\mbox{{\footnotesize d}}\rho}{\rho^5}\,\rho^{2 B+1}
\ln^E\frac{1}{\rho^2\Lambda^2}\int\limits_0^1\mbox{d}u\,\left(
\frac{2}{\sqrt{u\bar{u}}} K_1\left(\frac{\rho
m_\tau}{\sqrt{u\bar{u}}} \right) + \frac{\rho m_\tau}{u\bar{u}}
K_0\left( \frac{\rho m_\tau}{\sqrt{u\bar{u}}}\right)\right)\ee

\n with $\bar{u}=1-u$. The $u$-integration is sharply peaked at
$u=1/2$, which allows to put $u=1/2$ for examination of the
$\rho$-distribution. We plot the real part of this distribution
in Fig.2 as a function of $\alpha(m_\tau)$ and use
$\alpha(m_\tau)=(-\beta_0\ln m_\tau^2/\Lambda^2)^{-1}$ to relate
$\Lambda$ and $\alpha(m_\tau)$.

The distribution clearly shows two peaks. The first one is
located at values $\rho\sim (4-5)/m_\tau$, the precise coefficient
depending weakly on $E$ and $\alpha(m_\tau)$,
and corresponds to
the effective value of the instanton size $\rho_*\sim 400\,$ MeV.
The second one corresponds to the
integration region $\rho > 1/\Lambda$, and its contribution
should be small enough (say, by factor 5) in order that the
calculation makes sense. Combined with the requirement
that the value of the coupling at the effective scale is not
too large, say $\alpha(\rho_\ast) <1$, this restriction
yields the critical value of $\alpha(m_\tau)$, above which
the instanton contribution is ill-defined

\be \label{critical}
\alpha(m_\tau)_{cr}\approx 0.32\,,\ee

\n
 even if $\Pi^{ud}_{inst}$ is
still small. This value of the coupling lies well within the
interval under discussion  $\alpha(m_\tau) \sim 0.28-0.38$
\cite{ALE93}.
Note that
 since $\rho_* m_\tau\sim 4-5$ is a large number,
 the effect of including the
logarithms $\ln 1/(\rho^2\Lambda^2)$ into the integration
is important, and
yields a suppression by roughly a factor

\be \left(\frac{\ln (1/\rho_*^2\Lambda^2)}{\ln
(m_\tau^2/\Lambda^2)}\right)^E\,\sim\,\left(\frac{\alpha(m_\tau)}
{\alpha(\rho_*)}\right)^E\,\sim\,10^{-2}\ee

\n for typical values of $E$ and  $\alpha(m_\tau)$.
Technically, this effect comes from summation of higher-order
terms in the expansion in (\ref{projectorexpansion}).
\\


\n {\bf 3.}\hspace*{0.2cm}
We are now in the position to calculate the instanton
contribution to the $\tau$ decay widths. As usual, they are
normalized as

\be R_\tau\equiv\frac{\Gamma(\tau^-\rightarrow \nu_\tau+ \mbox{{
\footnotesize hadrons}})}{\Gamma(\tau^-\rightarrow\nu_\tau
e^-\bar{\nu}_e)}\,.\ee

\n Decomposing the correlation functions as

\be \Pi^{ij}_{\mu\nu,V/A}(q)=(q_\mu q_\nu-\delta_{\mu\nu} q^2)\,
\Pi^{ij}_{(1),V/A}(q^2)+ q_\mu q_\nu\,\Pi^{ij}_{(0),V/A}(q^2)\ee

\n and using analyticity in the cut $q^2$-plane, one can express
the total hadronic width $R_\tau$ as

\be\label{contourintegral}
R_\tau= 6\pi i\int\limits_{|s|=m_\tau^2}\frac{\mbox{
{\footnotesize d}}s}{m_\tau^2}\,\left(1-\frac{s}{m_\tau^2}
\right)^2\,\left[\left(1+2\frac{s}{m_\tau^2}\right)\,\Pi_{(1)}
(s)+\Pi_{(0)}(s)\right]\,,\ee

\n where $s=-q^2$ and

\be \Pi_{(J)}(s)\equiv |V_{ud}|^2 \left(\Pi^{ud}_{(J),V}(s)+
\Pi^{ud}_{(J),A}(s)\right)+|V_{us}|^2 \left(\Pi^{us}_{(J),V}(s)+
\Pi^{us}_{(J),A}(s)\right)\ee

\n with the CKM matrix elements $V_{ud},V_{us}$. Experimentally
the contribution from the strange current, $R_{\tau,S}$, can be
separated according to the net strangeness of the final state.
The nonstrange contribution can be further resolved into vector
and axial vector pieces, $R_{\tau,V}$ and $R_{\tau,A}$, according
to whether the final state contains an even or odd number of
pions. Thus, one defines

\be R_\tau=R_{\tau,V}+R_{\tau,A}+R_{\tau,S}\equiv
3\,\left(|V_{ud}|^2 + |V_{us}|^2\right)\,(1+\delta)\,,\ee
\be R_{\tau,V/A}\equiv \frac{3}{2}\,|V_{ud}|^2\,(1+\delta_{V/A})\,,
\ee

\n separating the ``Born'' term from the QCD corrections
$\delta$ (a small multiplicative electroweak correction is
understood), that are of the order of 20\%. We will now
determine the instanton contribution
$\delta^{inst},\delta^{inst}_{V/A}$ to the QCD corrections. In
the massless limit the correlation functions are transverse, see
eq.(\ref{instcorrelation}), i.e. $\Pi_{(0)}^{inst}(q^2)\equiv 0$.
The contour integral in (\ref{contourintegral}) is then taken
including logarithms of $s/\Lambda^2$ exactly, using the same
trick as above for the integration over the instanton size. This
produces immediately

\bea\label{tauwidth}
\delta_{V/A}^{inst}
\!&=&\! -\pi\cdot 4^B\left(\frac{2\pi^2}{3}\right)^{3}
(2\pi)^{6}\, c_1\,e^{-3 c_2+3 c_3}\,
\prod_{i=1}^{3}\left[\frac{\langle\bar{q}_i q_i
\rangle(\mu)}{m_\tau^3}\alpha(\mu)^{\gamma/3}\right]\,
e^{-2\pi/\alpha(m_\tau)} \\
&&\times\,\left(\frac{1}{\alpha(m_\tau)}\right)^{6+\gamma}
\,(-\beta_0\alpha(m_\tau))^E\, {\cal
P}_E^\epsilon\,\Bigg\{e^{-\epsilon/(\beta_0\alpha(q))}
\,F(\epsilon)\,H(\epsilon) \Bigg[\frac{1}{2
B-1-2\epsilon}\pm 1 \Bigg]\Bigg\}\,,\nonumber\eea

\n which is our final result. The upper sign holds for vector
currents, the lower sign for axial vector currents. The effect
of the contour integration -- up to constant factors --  is
contained in the function $H(\epsilon)$,

\be H(\epsilon)=\sin\,\pi (B-\epsilon)\,\left[\frac{1}{1-B+
\epsilon}-\frac{3}{3-B-\epsilon}+\frac{2}{4-B+\epsilon}\right]\,.
\ee

\n Note that the expansion of $H(\epsilon)$ starts with order
$\epsilon$, since $B$ is integer. Therefore the transition from
the correlation functions to the decay widths suppresses the
instanton contribution by one power of $\alpha(m_\tau)$.
With $\langle\bar{s} s\rangle = \langle\bar{u} u\rangle$ the
instanton contribution to the strange decays equals that to the
nonstrange decays. Then

\be \delta^{inst}=\frac{1}{2}\,(\delta_V^{inst}+\delta_A^{inst})
\simeq \frac{1}{20}\,\delta^{inst}_V\,.\ee

\n The term differing in sign for the vector and axial vector
widths in (\ref{tauwidth}) cancels in the total width, leaving
only the first term with the small coefficient $\sim 1/(2
B-1)$. Thus there is a strong cancellation in the sum of vector
and axial vector contributions, which is similar in effect  to the
 cancellation in the contribution from four-quark
operators in the standard SVZ expansion \cite{BRA92}. However,
this cancellation relies strongly on the factorization
approximation, which eliminates the second and third term in
(\ref{ope}) having much larger coefficients. For this reason,
the instanton contribution to the total width is much more
sensitive to deviations from factorization and therefore less
reliable numerically. Since
   $\delta^{inst}_V=-\delta^{inst}_A+2\delta^{inst}\,,$
the axial vector contribution is essentially of the same
magnitude as for the vector current but with an opposite sign.
In Fig.3 we plot
$\delta^{inst}_V$ and $\delta^{inst}$ as a function of
$\alpha(m_\tau)$ for three choices of the anomalous dimension
of the $(\bar q q)^3$ operator $\gamma$ = 2/3, 4/3, and 6/3,
corresponding to
 $E$ =28/9, 34/9, and 40/9, respectively. (The middle value
corresponds to factorization.)
At $\alpha(m_\tau) = 0.32$ we get

\be \label{numbers}
\delta^{inst}_V\simeq -\delta^{inst}_A =
0.03 - 0.05\,,\hspace{0.5cm}
 \delta^{inst} \simeq 0.002 - 0.003\,.
\ee

\n At larger values of the coupling $\alpha(m_\tau) \sim 0.34-0.36$
the instanton contribution blows up, but our calculation is no
longer under control.
Thus the instanton contribution is essentially of the same size
as the contribution from dimension-6 operators
$\sim~\langle \bar q q\rangle^2/q^6$ \cite{BRA92},

\be \delta^{D=6}_V = 0.024\pm 0.013\,,
\hspace{0.5cm}\delta^{D=6}_A =-(0.038\pm 0.020)\,,
 \hspace{0.5cm}\delta^{D=6} = -(0.007\pm 0.004) \,,\ee

\n which are the largest power corrections in the standard approach.
\\



\n {\bf 4.}\hspace*{0.2cm}
We performed an explicit calculation of the instanton
contribution to the $\tau$-decay widths, which corresponds
to the exponential correction to the coefficient function
in front of the six-quark operators in the operator
product expansion of the relevant correlation functions, and
which is distinguished by the
chiral properties of the instanton-induced
effective vertex. Contrary to previous claims, we find
a sizable
instanton contribution, which for $\alpha(m_\tau)\approx 0.32$ is
of the same order
of magnitude as standard nonperturbative corrections
 induced by nonzero vacuum expectation values of the
operators of lowest dimension. An inherent uncertainty of
our calculation is about a factor of two for vector and axial
vector channels, and maybe larger for the total width owing to
strong cancellations. For still higher values of $\alpha(m_\tau)$
we cannot make quantitative statements, but there is no reason
to expect that the instanton contribution becomes small.

 Unfortunately, as far as numerical values are concerned,
our result is of
limited practical importance, since the effective instanton
size has proved to be large, $\rho_\ast\sim 400\,$MeV.
At such low scales, the instanton density is modified
strongly by external gluon fields. The effect is conceptually
quite similar to the effect of the quark condensate in
(\ref{effmass}), and the lowest-order correction reads \cite{SHI80}

\be \label{effdense}
d_{eff}(\rho) = d(\rho)\left[
1+ \frac{\pi^4\rho^4}{8\alpha^2(\rho)}
\left\langle \frac{\alpha}{\pi}G^2\right\rangle\right]\,,
\ee

\n where $\langle\alpha/\pi\,G^2\rangle\simeq 0.012
\,\mbox{GeV}^4$
is the gluon condensate. With the effective instanton size 400 MeV,
the correction term in brackets in (\ref{effdense}) is several
times larger than unity, indicating that the effect is important.
For the case of $\tau$-decay, this means that instanton
contributions to the coefficient functions of operators of
higher dimension, including gluon fields in addition to six quark
fields, are likely to be larger than the
one considered in this paper, so that the expansion in
powers of $1/m_\tau$ fails. One may speculate
that this expansion is at best an asymptotical one, and
the contributions $\sim 1/m_\tau^{18}$ under discussion
are already in the region where the series starts to diverge.
Anyhow, in agreement with the old wisdom \cite{SHI80}, we
find that instanton calculations are hardly useful for
quantitative estimates of nonperturbative effects, but
rather can be used to indicate a scale at which the power
expansion breaks down.

Thus, our conclusion is partly pessimistic.
On the one hand,
we find it difficult to  justify the use of the operator
product expansion in finite energy sum rules at the scale
of the $\tau$ mass, since contributions of higher orders are of
the same order as leading power corrections. Our
calculation essentially supports the old philosophy of SVZ
\cite{SHI79}, who introduced Borel sum rules in
which higher-order power corrections are suppressed by
factorials. In our case Borel transformation would
introduce a factor $1/8!\simeq 2.5 \cdot 10^{-5}$ and render the
instanton contribution completely negligible.
In the practice of the numerous QCD sum rule calculations
this old argumentation has partly been forgotten. The reason is
that higher-order corrections have never been calculated
explicitly, and as far as leading-order power
corrections are concerned,
Borel and finite energy sum rules give very similar
results, within the typical 10\% accuracy.
It is the requirement of a very high precision,
which makes mandatory to reconsider the theoretical accuracy
of the sum rule program for $\tau$-decays. In the light of
our result a simultaneous extraction of $\alpha(m_\tau)$ and
nonperturbative parameters from $\tau$ decays \cite{ALE93}
is hard to justify.

On the other hand,
all nonperturbative contributions to the total $\tau$-decay
width prove to be small, less or
of order 1\%. Although the leading
power corrections cannot be guaranteed to improve the accuracy,
they can
 well  indicate an intrinsic uncertainty
of the theoretical prediction. Thus, it is plausible to expect
an accuracy of
the perturbative prediction for the $\tau$ hadronic width
of order 1\%, but the situation could be substantially worse
for the exclusive vector and axial vector channels.
This allows for a determination
of the QCD coupling within 10\% accuracy, which is competitive
to the current accuracy of the determination of $\alpha$ at the scale
of the $Z$-boson mass, but,
in difference to the latter, cannot be substantially improved.
\\


\n {\sc Acknowledgements.}\hspace*{0.2cm}
The work by I.B. was supported by the US Department of Energy
under the grant DE-FG02-90ER-40577.




\newpage
\normalsize
\n {\large\bf Figure Captions}\\[0.3cm]
\newcounter{captionlist}
\begin{list}{\bf Fig.\arabic{captionlist}} {\usecounter{captionlist}}

\item The four types of diagrams that contribute to the Wilson
coefficient of six-quark operators. The instanton is depicted as a
$2 n_f$-quark vertex with only the flavours $u$ and $d$ shown explicitly.
A circled cross denotes the insertion of the current.

\item Distribution of instanton sizes as a function of
$\alpha(m_\tau)$. The vertical scale is arbitrary.

\item Instanton contribution to the vector decay channel (a) and
the total width (b) for three values of ``anomalous dimensions'',
$E=28/9\,,34/9\,,40/9$, as a function of $\alpha(m_\tau)$.
\end{list}


\end{document}